# Toward the Explainability of Protein Language Models

Andrea Hunklinger[1,2], Noelia Ferruz[1,3,*]

[1]Centre for Genomic Regulation, the Barcelona Institute of Science and Technology, Dr Aiguader 88, Barcelona 08003, Spain
[2]Universitat de Barcelona, Facultat de Farmàcia I Ciències de l'Alimentació, Avda. Diagonal 643, Barcelona 08028, Spain
[3]Universitat Pompeu Fabra, Barcelona, Spain
*E-mail: noelia.ferruz@crg.eu

**Protein language models (pLMs) excel in a variety of tasks that range from structure prediction to the design of functional enzymes. However, these models operate as *black boxes*, and their underlying working principles remain unclear. Here, we survey emerging applications of explainable artificial intelligence (XAI) to pLMs and describe the potential of XAI in protein research. We divide the workflow of protein AI modeling into four information contexts: (i) training sequences, (ii) input prompt, (iii) model architecture, and (iv) input-output pairs. For each, we describe existing methods and applications of XAI. Additionally, from published studies we distil five (potential) roles that XAI can play in protein research: Evaluator, Multitasker, Engineer, Coach, and Teacher, with the Evaluator role being the only one widely adopted so far. These roles aim to help both protein scientists and model developers understand the possibilities and limitations of implementing XAI for predictive and generative tasks. While our analysis focuses on pLMs, both this categorization and roles are broadly applicable to any other model architectures. We conclude by highlighting critical areas of application for the future, including risks related to security, trustworthiness, and bias, and we call for community benchmarks, open-source tooling, domain-specific visualizations, and wet-lab characterization to advance the interpretability of protein AI.**

**Glossary**

**Ante-hoc and post-hoc interpretability:** Approaches to XAI where interpretability is achieved by model design (ante-hoc) or through analysis after training (post-hoc).

**Autoregressive model:** A modeling objective where a model predicts the next element in a sequence based solely on preceding elements.

**Contrastive explanation:** An explanation that identifies the features responsible for why a model produced one outcome instead of another specific alternative.

**Counterfactual explanation:** An explanation that describes the minimal change to an input that would cause the model to produce a different outcome.

**Explainable artificial intelligence (XAI):** A research field focused on making the decisions and internal mechanisms of black-box machine learning models humanly understandable.

**Feature attribution:** Assessment of how much a specific part of the input (a feature) contributed to the generation of the output. In protein AI modelling, depending on the tokenizer used, a feature may correspond to a single amino acid or a short sequence.

**Layer-wise relevance propagation (LRP):** A feature attribution method that propagates the output relevance backwards through the model to the input layer to determine the contribution of each input feature.

**Local Interpretable Model-Agnostic Explanations (LIME):** A feature attribution method that approximates a complex model locally with an interpretable surrogate model to explain individual predictions.

**Mechanistic interpretability:** A subfield of XAI research focused on understanding the internal workings of models by identifying specific components and analyzing how they implement computations or circuits.

**Model pruning:** Strategic reduction of model parameters to improve efficiency, typically by removing weights or neurons with minimal impact on performance.
**Natural language processing (NLP):** The application of machine learning to the analysis and generation of human language, such as English, enabling tasks like translation, summarization, and question answering.
**Neuron:** A computational unit within a model component, such as a neural network layer or a multi-head attention head, responsible for processing and transforming input data.
**Prompt injection:** An attack method in which inputs are deliberately crafted to override a model's intended instructions or induce unintended behavior, potentially compromising the reliability or safety of generative AI systems.
**Protein language model (pLM):** A neural network trained to estimate the probability (likelihood) of amino-acid sequences -analogous to text language models- thereby capturing statistical patterns that can be exploited for predicting structure, inferring function, and guiding protein design.
**Residual stream:** The intermediate data representation passed through layers of a Transformer model.
**Shapley Additive Explanations (SHAP):** A feature attribution method that approximates Shapley values for machine learning models by sampling subsets of features and measuring input–output changes to estimate each feature's contribution to the prediction.
**Shapley values:** A mathematical concept from cooperative game theory in which each feature's value is its average marginal contribution to the outcome computed across all possible subsets of features.

## Introduction

Significant advances in computational hardware and deep-learning software have enabled the adoption of natural language processing (NLP) models in everyday applications.[1] In parallel, the decreasing cost of DNA sequencing and the extension of protein structure databases have enabled the training of ever-larger protein language models (pLMs). pLMs now set the state of the art in tasks as diverse as the prediction of drug-target interactions[2], protein structure prediction[3,4], and protein design[5]. Despite their applicability, these models often function as "black boxes", making it difficult to interpret their decision-making processes. This lack of transparency undermines trust in their predictions and raises important concerns regarding their safe and responsible deployment, particularly when simpler, inherently interpretable models fall short of the accuracy demanded by modern applications.[6]

In response to these challenges, the field of Explainable Artificial Intelligence (XAI) has gained traction in the last years. XAI aims to enhance the transparency of machine learning (ML) models by approximating their internal reasoning or by visualizing the patterns they learn from data. These approaches help bridge the gap between model complexity and human interpretability;[6] however, applying them to biomolecular language models remains technically demanding.

Here, we survey the emerging intersection between XAI and pLMs. We categorize XAI applications based on the origin of the information within the modeling workflow - namely, the training dataset, the input query, the model architecture, and input-output pairs (Fig. 1). Later, we analyze the intended purposes of XAI in prior studies and define five potential roles - Evaluator, Multitasker, Engineer, Coach, and Teacher - that explainability methods could play within a workflow. Although our primary focus is on pLMs, most commonly trained with the Transformer architecture[7], the proposed framework is readily applicable to other architectures used in protein research, including diffusion models, graph neural networks (GNNs), or AlphaFold - and we highlight relevant XAI applications for these cases throughout the manuscript. For readers interested in the taxonomy and technical details of current XAI methods in natural language processing (NLP) and computer vision (CV), several review articles are available in the literature.[8–18] We conclude by outlining future directions, including the need for expanded benchmarking efforts, the need for wet-lab characterizations, the development of domain-specific explainability and visualization tools, and a critical examination of risks related to security, trustworthiness, and bias. Collectively, our analysis aims to describe the emerging applications of the rapidly evolving XAI field and promote its establishment as a tool for the discovery of biological principles.

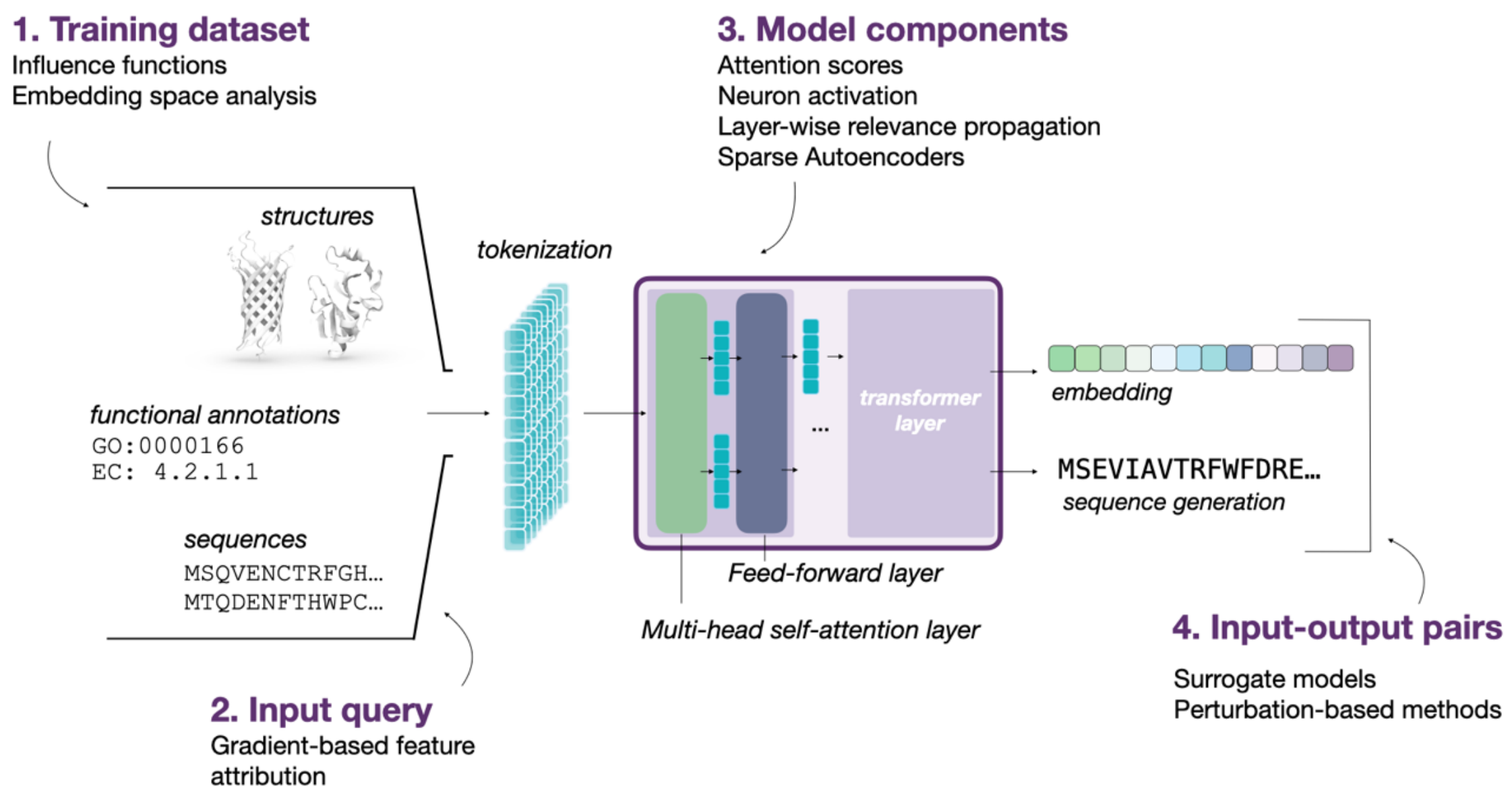

***Figure 1: Conceptual overview of XAI methods across the protein modeling workflow.*** *XAI approaches can be grouped according to their source of information: (1) training dataset (influence functions, embedding analyses), (2) input query (gradient-based feature attribution), (3) model components (attention scores, layer-wise relevance propagation, neuron activations, sparse autoencoders), and (4) input–output pairs (surrogate models and perturbation-based methods). The schema illustrates these categories using a Transformer-based pLM that incorporates multiple modalities (sequences, functional annotations, and protein structures), but the taxonomy can be applied to other model architectures and input types. Categories are numbered by order of discussion in the text.*

### Training dataset: Influence functions expose dataset biases

PLMs are trained on large datasets of protein data. While originally trained on large corpora of sequences[19,20], most recent paradigms often incorporate structural information[21–23] and functional annotations[22,24,25] (Fig. 1). Because these heterogeneous data types must be compatible with the mathematical operations of the model, each modality requires an appropriate tokenization. Functional annotations are typically encoded as textual tokens or categorical labels[24,25], whereas structural information is transformed in several ways: via predefined structural alphabets such as 3Di (e.g., in ProstT5[23]), through fused sequence-structure (as in SaProt[21]), or through learned structural representations obtained with vector-quantized autoencoders, as in ESM-3[22]. Importantly, while tokenization determines how information is processed, it does not eliminate the biases already present in the underlying biological data, i.e., systematic patterns or overrepresentation of specific traits within the body of known proteins, which may include sequence composition bias,[26,27] uneven sampling across species[28] or geographic regions[29,30], and technical biases arising from protocol-dependent variations.[31,32] Biases in sequence data are also reflected in the ways models capture evolutionary information from the training sets.[28] Before pLMs, often most high-performing approaches used Multiple Sequence Alignment (MSAs) for various tasks, from remote homology detection[33], to structure prediction. pLMs were originally proposed as single-sequence alternatives that could compensate for sequences with few homologs[34,35]. However, early community assessments showed that pLM-only approaches do not surpass MSA-based pipelines in many tasks, and the performance of pLM-based models is worse for proteins that do not have many homologs in the sequence databases[36]. In CASP15 (2022), MSA methods remained dominant for structure prediction.[36,37] Similarly, on ProteinGym, a large benchmark on protein fitness variant prediction, pLM performance showed to correlate with MSA depth for the assayed protein family[38].

While ameliorated via model and data scaling, this dependency on homologous data can become counterproductive. Recently, Gordon et al.[39] demonstrated that pLM success on fitness prediction correlates with the log-likelihood of the wild-type sequence: both over-preferred and under-preferred wild-type sequences negatively impact the model's performance. The authors suggested performing unsupervised fine-tuning only in regions of low-likelihood space. Performing them on high-likelihood regions can, in fact, worsen performance, a finding also confirmed by Hsu et al.[40] Next, they generalized the effect of training sequences from the species level to individual sequence probabilities. In particular, they applied *influence functions* to analyze how specific protein sequences impact the performance and generalization of pLMs.

Influence functions estimate how a model's predictions change when training examples are perturbed or removed, tracing behavior back to data points to diagnose biases (**Fig. 2a**). Mathematically, influence functions measure the first-order change in an individual sequence's loss under the model when it is infinitesimally upweighted, quantifying how much that sequence shapes the model's prediction. First introduced in the field of robust statistics in 1974[41], they have only recently been applied to deep learning[42] and large language models (LLMs)[43], thanks to approximations that reduced their computational cost. Gordon et al.[39] found that applying influence functions to pLMs revealed that the distribution of influential data points follows a power-law distribution, with homologous protein training data being the most influential.

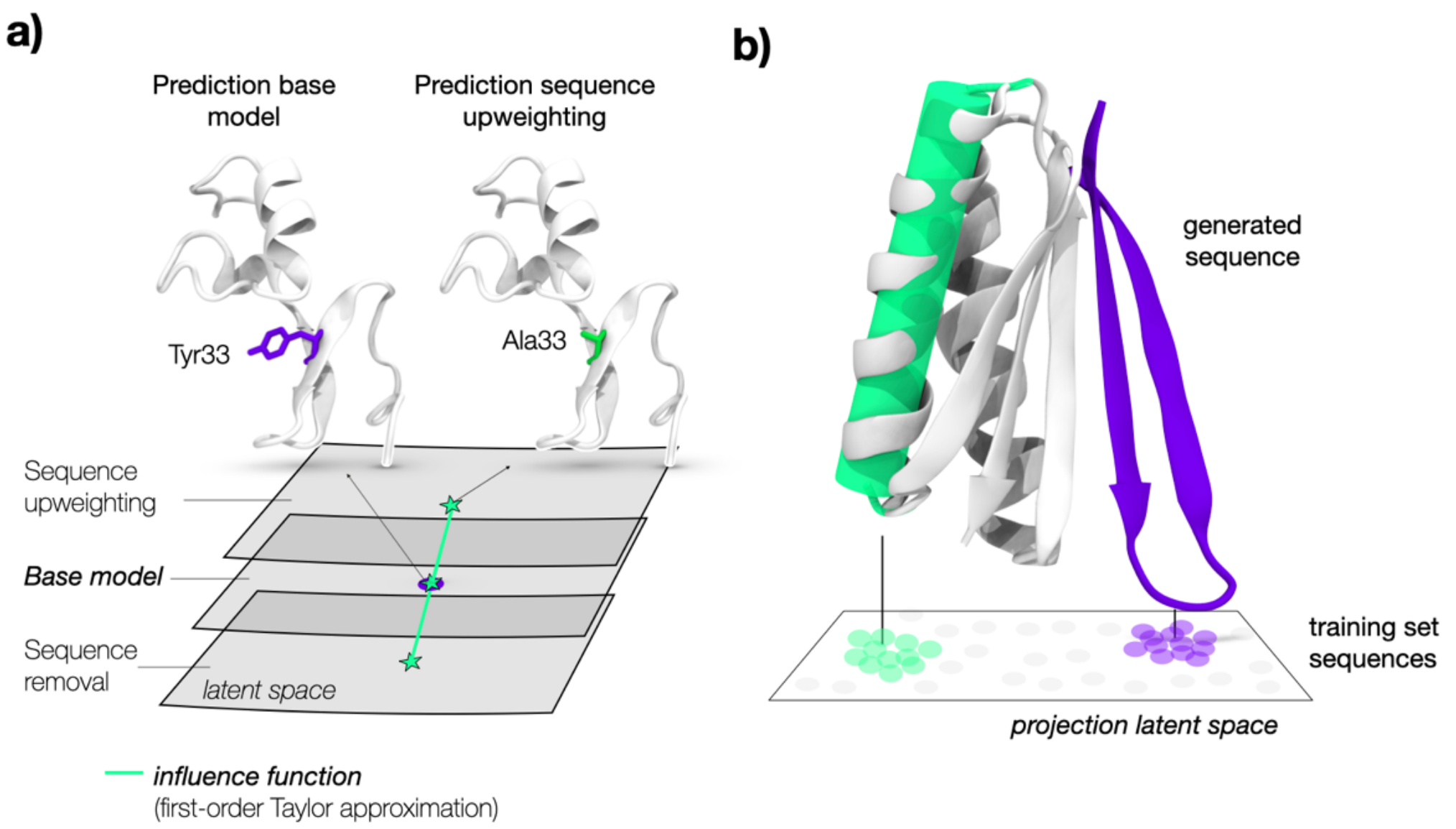

***Figure 2**: **Explainability through analysis of training data. (a)** Influence functions estimate how upweighting or removing individual training sequences affects model outputs and parameters, using a first-order Taylor approximation of the loss around the base model. Removing a training set sequence can alter the model during inference, for example, by predicting a different amino acid. **(b)** Predicted and training sequences can be projected as embeddings in a latent space, where their similarity (computed through distances such as cosine similarity) reveals influential or outlier examples and helps assess relationships between generated and training sequences.*

In addition to influence functions, a complementary strategy is to project both training and generated sequences into the model's latent space and measure embedding distances directly (Fig. 2b). Littmann et al. used this strategy to predict functional annotations, like Gene Ontology terms, through transfer based on the proximity of proteins in the SeqVec embedding space.[44] In the NLP realm, exBERT[45] provides a visualization technique for LLMs that allows users to inspect the closest neighbors in embedding space for each generated token in comparison to an annotated, smaller dataset. The same interface overlays attention maps, providing a multi-scale view from dataset bias to token-level rationale. Parallel efforts, such as studies on the effects of data leakage[46] or dataset composition[47], could be combined with XAI insights to advance understanding of how decisions made prior to training shape model behavior.

**Input query: Gradient-based feature-attribution methods reveal key residue networks**

Gradient-based feature-attribution methods (Box 1, Fig. 3) quantify how each input feature influences a model's output. Because gradients are available via the standard backward pass, explanations can be extracted post-training with no architectural changes. Several variants have been developed over the years. Some, like Gradient × Input (Eq. 1), directly use the gradients of the model's output $f(x)$ with respect to the *i*-th input feature $x_i$, while others, like Integrated Gradients (Eq. 2), accumulate gradients along a path from a baseline $x'$ to the input, using an interpolation scalar $\alpha$. We refer readers to the review of Wang et al.[48] for a more in-depth technical overview.

$$attribution_i = \frac{\partial f(x)}{\partial x_i} x_i \tag{1}$$

$$attribution_i = (x_i - x_i') \int_0^1 \frac{\partial f\left(\alpha x + (1-\alpha) x'\right)}{\partial x_i} d\alpha \tag{2}$$

Gradient-based feature attribution methods have been applied to pLMs in tasks such as the prediction of disordered regions[49] and antibody affinity[50] (Fig. 3a). In each study, token-wise attribution scores were mapped back onto the input sequence and compared with established biophysical heuristics, like for example, hydrophilicity predictions. The strong agreement led the authors to conclude the model was able to identify the necessary features and distinguish informative from non-informative residues.

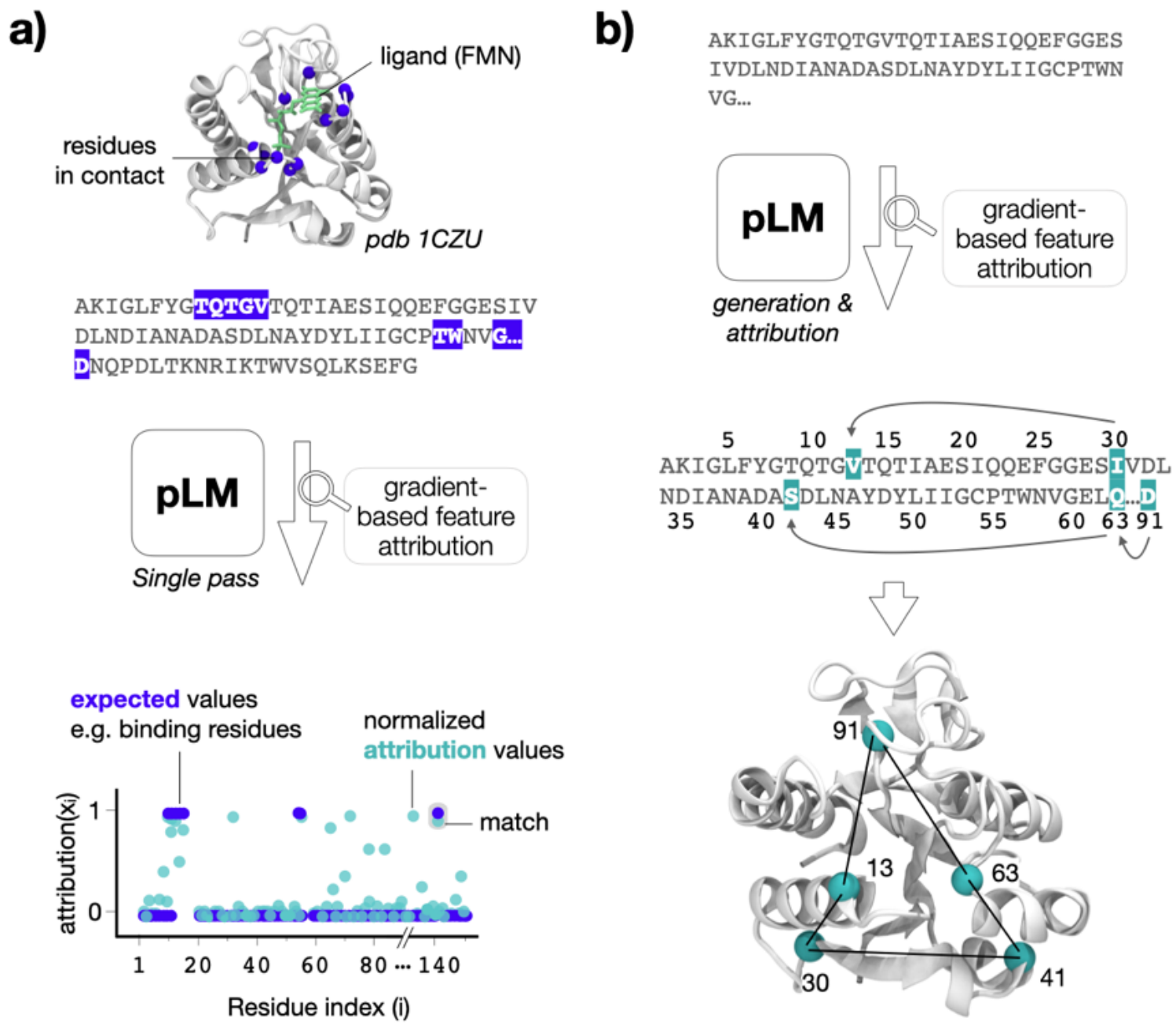

***Figure 3**: **Explainability through analysis of input sequences. (a)*** *Gradient-based feature attribution methods can produce per-residue ($x_i$) attribution scores for a given sequence, respective to a model's prediction of a property, in this case whether residues are in contact (violet) with a ligand (FMN). These computed attributions (lower plot, blue) can be compared with the experimentally measured residues (violet).* ***(b)*** *In a different scenario, attributions can also be computed for the features (residues) that most impact the autoregressive generation of the next token. When done for each position in the protein, this process could reveal networks of amino acids that connect to one another (e.g., shown residues 13, 30, 41, 63, and 91).*

## BOX 1: Feature attribution methods

In ML, a *feature* is an individual, measurable characteristic of the data used to make predictions. For example, in a housing price prediction model, features might include square footage, number of bedrooms, or location. In the context of pLMs, features correspond to the input tokens - most often single amino acids. *Feature attribution* techniques are a family of XAI methods that aim to explain which features most influence the model's output, helping to interpret and understand the model's behavior. They are commonly divided into four categories: **gradient-based, decomposition-based, perturbation-based, and surrogate-based.**[13] Although all approaches ultimately produce the same type of output - an attribution value for each input feature - they rely on different strategies. In our information-context categorization, they therefore belong to different groups: input query, model components, and, for the latter two, input–output pairs.

*Gradient-based* feature attribution methods estimate importance by computing how small changes in the input affect the model's output, typically via gradients or gradient-integral paths, to explain, "Given the current prediction, which input positions effected the greatest influence?"[48]. *Decomposition-based* methods, exemplified by Layer-wise Relevance Propagation (LRP), assign relevance scores by decomposing the model's prediction across its internal components.[51] These two classes require access to model internal information, like gradients, activations and layer-wise computations, and have sometimes been described as "*white-box*" approaches, as they rely on internal signals[52]. In contrast, the next two categories treat the model as a black box and infer attribution from probing input-output behaviors, without having access to the network internal workings. *Perturbation-based* methods assess feature importance by systematically altering parts of the input and observing the resulting changes in the output. In addition to occlusion and mutagenesis studies (systematically mutating input residues and observing output changes), Shapley Additive Explanations (SHAP)[53] is one of the most prominent perturbation-based techniques and it approximates the contribution of each feature by sampling from all possible feature combinations. Local Interpretable Model-Agnostic Explanations (LIME)[54], on the other hand, is a *surrogate-based* method that creates an interpretable model to approximate the original function of the black-box model. These probing approaches are generally better at identifying the necessary and/or sufficient features for a specific change in the output compared to gradient-based methods, which focus more on individual influential positions.[55]

Improving the faithfulness, robustness, and fairness of explanation methods - particularly feature attribution techniques - is an active area of research.[56] Even when a model identifies biologically relevant patterns and employs them for output generation, XAI methods may fail to accurately reflect this internal process. Each method relies on its own explanation strategy, which can introduce approximations and deviate from the model's true behavior. As a result, different methods may produce conflicting signals, a challenge in the absence of ground-truth data.[57,58] In practice, it is advisable to assess faithfulness using fidelity frameworks and to benchmark multiple approaches[59].

The application of gradient-based attribution methods has still unexplored potential in protein research. For instance, such methods could identify learned and unwanted biases from the training data, as seen in the "Clever Hans" example in computer vision, where a tag in the corner of the input image influenced the prediction more than the actual content of the image.[60] In the case of generative models, the generation of the next amino acid could then be viewed as a prediction task, in which the model selects one of the *n* possible amino acids. Extending this across an entire sequence could reveal networks of co-evolving or allosterically connected residues (**Fig. 3b**). Multi-modal approaches that integrate protein sequences with structural information and functional annotations, whether through separate tokenization[23,61] or structure-aware sequence tokens[21], represent compelling targets for explainability research. Such models may provide more direct associations among residues, structural features and potentially functional determinants, but we are not aware of any publications to date. Additionally, relevant patterns in the input, such as for example, catalytic residues in an enzyme, could also be integrated in the loss function, rewarding attribution on biologically meaningful patterns while penalizing unwanted biases. This approach has been shown to be successful in computer vision tasks,[62,63] suggesting potential to be transferred to pLMs.

Despite these prospects, there are some limitations. Over the years, numerous gradient-based feature attribution methods have been developed, and identifying the most reliable one for a given task is nontrivial. Often, new techniques have been introduced to address the issues of older methods, such as the introduction of the *sensitivity axiom* for the Integrated Gradients method by Sundararajan et al.[64] However, this method requires a baseline input, and selecting an appropriate baseline value can be crucial.[65] Research has also cautioned against applying common gradient-based attribution methods designed for neural networks directly to Transformer architectures, as they often fail to capture the full complexity of these models.[66] Consequently, it is advisable to test multiple methods and compare their results[59]. Gradient-based methods also face challenges with the discrete inputs used in language models, because distances between discrete tokens are not straightforward to compute. As a result, methods such as Integrated Gradients are usually applied to the token embeddings[64], a model-internal continuous representation of the input. Alternatively, further efforts are made to adjust these methods for use in discrete spaces,[67,68] but they have not yet been tested on pLMs.

**Model components: single neurons, attention scores, and sparse autoencoders**

PLMs have been most commonly trained with the Transformer architecture.[69] **Fig. 4** outlines the key components of an encoder-only Transformer; for an in-depth analysis of the architecture, we refer the readers to recent work.[7,70] The Transformer is composed of multi-head attention blocks (Fig. 4), which allow the model to focus on different parts of the input sequence and capture long-range dependencies independent of position. Because *attention scores* are easy to inspect, they have become a popular target for interpreting Transformer architectures. In particular, attention score analysis is often used for internal retrospective validation:[71–77] researchers search for patterns related to specific biological properties in the map of high and low attention scores within particular layers and heads of the multi-head attention module. When such patterns appear, it is often concluded that the model has captured this biological aspect by encoding it in its attention weights. For instance, Koyama et al.[78] used a TCR binding predictor model and identified that highly attended residues overlapped with structural properties such as hydrogen bonds and residue distances, while Kannan et al.[79] used the attention scores between spatially distant residues to predict allosteric sites.

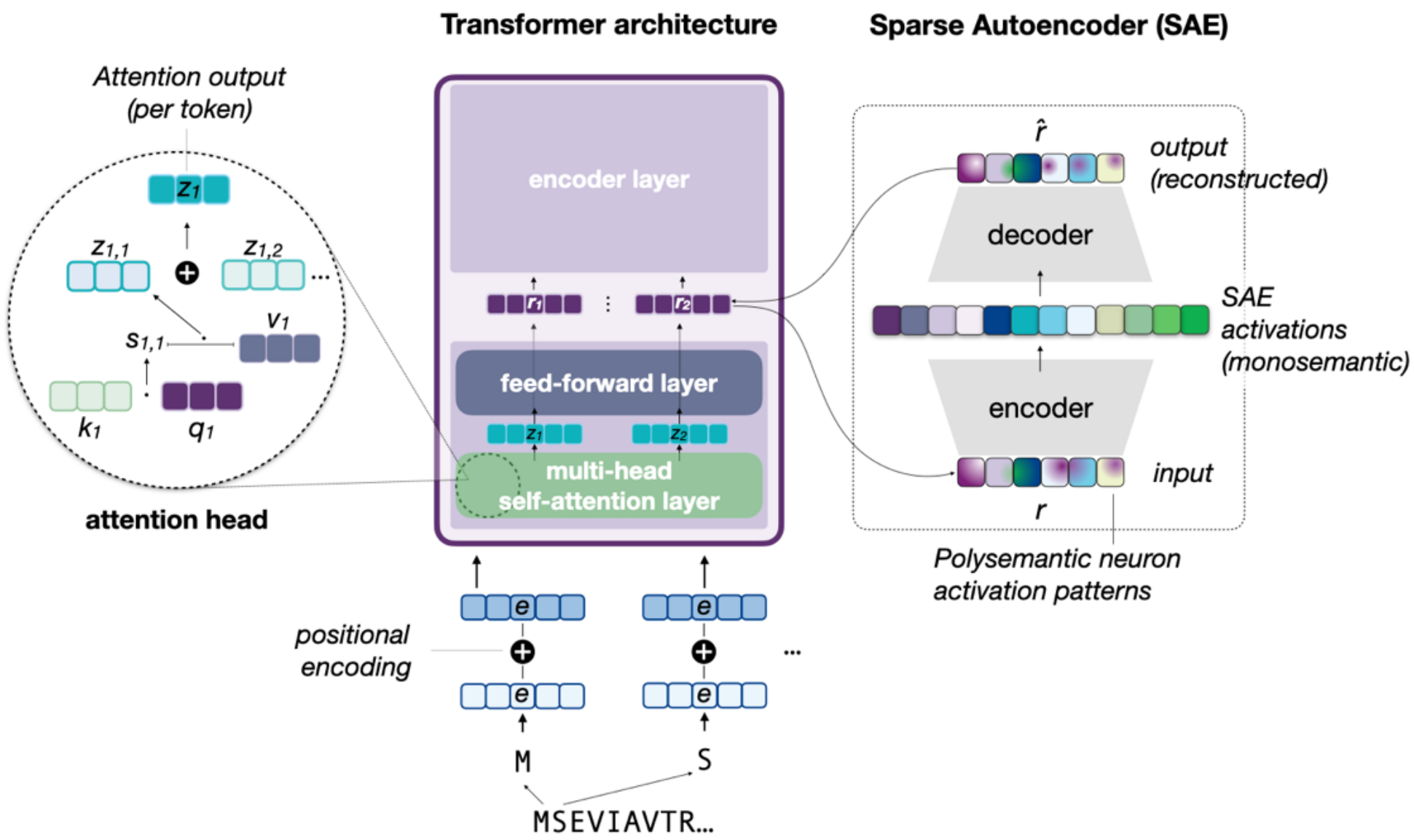

***Figure 4**: **Overview of the Transformer architecture and its analysis with XAI methods.*** *Each input token (e.g., amino acids, M, S…) is converted into an embedding vector (e) combined with positional information before being processed through successive encoder layers. Within each attention head, query ($q_i$), key ($k_i$), and value ($v_i$) vectors produce attention outputs ($z_i$), which represent weighted combinations of contextual information across tokens. The attention output and the feed-forward layer output contribute to the residual stream connecting layers ($r_i$). An SAE trained on these activations can disentangle polysemantic neuron responses into monosemantic features, enabling interpretation of how information is represented, localized, and propagated through the model.*

In NLP, when individual heads consistently capture specific patterns in text, they are termed *specialized heads*.[80] Voita et al.[81] identified positional, syntactic heads, and rare-word heads in an encoder-decoder Transformer. For pLMs, Vig et al.[82] analyzed the correlation between attention scores and ground-truth annotations of contact maps, binding sites, and post-translational modifications (PTMs) across all heads and layers. They found that the attention maps aligned most strongly with contact maps in the deepest layers, with binding sites across most layers of the models, and with PTMs in only a small number of heads. Wenzel et al.[83] further isolated the influence of individual heads by subtracting their latent representations. Similar specialized heads have been observed in DNA[84,85] and RNA[86] language models.

Despite these advances, insights obtained via attention score analyses have yet to be leveraged for efficiency gains. Indeed, the actual contribution of the attention mechanism in generating the output remains debated.[87,88] Voita et al. conducted model pruning experiments (a strategic reduction of model parameters[89,90]) and showed that non-specialized heads, particularly those in the encoder self-attention mechanism, can be removed without significantly affecting the model's performance.[81] Automatic frameworks for the detection of redundant weights have also arisen, such as the LLM Pruner[91]. Yom Din et al.[92] demonstrated that removing 7.9% of GPT-2 retained 95% accuracy, and other studies reported that the attention mechanism accounts for only 30% of the BERT embeddings,[93] with negligible performance loss when attention matrices are removed.[94] This raises the question of whether the analysis of the attention module can explain the model behavior. Even if attention maps reflect biologically relevant interactions, it is not guaranteed that this information is used for generation or prediction.

A different approach for model component analysis consists of using *Layer-wise Relevance Propagation* (LRP).[51] LRP is a decomposition-based feature attribution method (**Box 1**) initially developed for convolutional neural networks (CNNs) in image classification, where relevance scores for each neuron are computed by starting from the final output layer and moving backward toward the input layer. LRP has been applied in biology for graph-based neural networks[95,96] and fully-connected neural networks[97]. Efforts have been made to extend its use to Transformers for NLP tasks.[98] Achtibat et al.[99] recently introduced AttnLRP, where they customized the propagation rules to account for the various components in Transformer models. They identified the most relevant neurons for various NLP concepts and were able to influence the generation by up- or down-regulating the activation of these neurons. LRP could be applied to Transformer-based pLMs to evaluate the importance of features or even to influence the generation process.

Researchers in NLP have also studied the effect of individual neurons, revealing that lower layers capture syntactic information, while higher layers encode abstract contextual relationships.[100] By observing when a single neuron or a group of neurons activates, researchers can infer which concept the neuron has learned and its role in generating the model's output.[101,102] However, neurons are often activated for multiple, unrelated concepts - a phenomenon known as polysemanticity (**Fig. 4**). In this way, features are stored in superposition, meaning a neuron's activation can represent a combination of multiple signals. This allows the model to express more concepts than there are neurons, enabling greater efficiency and flexibility,[103] at the expense, however, of interpretability. To avoid this limitation, a technique that is gaining considerable attention in recent years are sparse autoencoders (SAEs). SAEs attempt to capture the information of highly complex, polysemantic neuron units in standard models into sparse, monosemantic neurons that encode specific interpretable features.

Mechanistic interpretability aims to understand how model components and circuits contribute to the model's decision-making processes.[104] SAEs[105] are emerging as a powerful technique in this field: they take embeddings or residual stream activations, learn a higher-dimensional latent representation, and use a decoder component to reconstruct the original input (**Fig. 4**). A sparsity constraint ensures that only a small fraction of the neurons activate simultaneously, encouraging specialization and mitigating polysemanticity (Fig. 4).[106]

SAEs have recently been applied to pLMs embeddings. Simon et al. introduced an SAE called InterPLM[107] and additionally implemented an automated annotation system using a natural language model. With their model called InterProt, Adams et al.[108] found that protein family-specific features were most prominent in the early-to-mid layers and even declined in the later layers, which demonstrated the importance of choosing the right representation layer for different prediction tasks, while Gujral et al.[109] found many features that were associated with specific protein families. SAEs were also applied to the DNA language model Evo2, where the latent features correlated with elements such as exons, introns, and transcription factor motifs.[110,111] In the realm of protein design, Parsan et al.[112] not only trained an SAE to explain the pLM but also used it to influence the downstream structure prediction model, ESMFold[113], through a technique called feature steering[114]. They identified a feature strongly correlated with residue hydrophobicity, overactivated it, and used the modified reconstruction of the embedding as input for ESMFold. Recently, Boxó et al. identified features that correlate with enzyme activity, and used them to steer the pLM ZymCTRL[25] for the generation of more active alpha amylases[115].

In addition to SAEs, pLM embeddings have been widely used for a variety of tasks.[116] For instance, ESMFold[113] leverages such embeddings for structure prediction without requiring MSAs. In the context of XAI, Li et al.[117] showed that the size of the embedding model particularly impacts performance on tasks dependent on coevolutionary signals, with deeper models capturing these patterns more effectively. However, for fitness prediction tasks, longer pretraining did not improve transfer learning performance, and the authors argued the same could apply to generating artificial proteins with low sequence identity to natural proteins, as these tasks do not rely on learned coevolutionary signals. They showed that in such cases, the performance is driven by features learned in the earlier layers. Overall, predicting a protein structure does not guarantee understanding of protein biology or having acquired physical or chemical knowledge.[118] This reiterates previous findings in attention scores that correlation does not imply use, and it motivates validating embedding features with targeted interventions.

### Input-output pairs

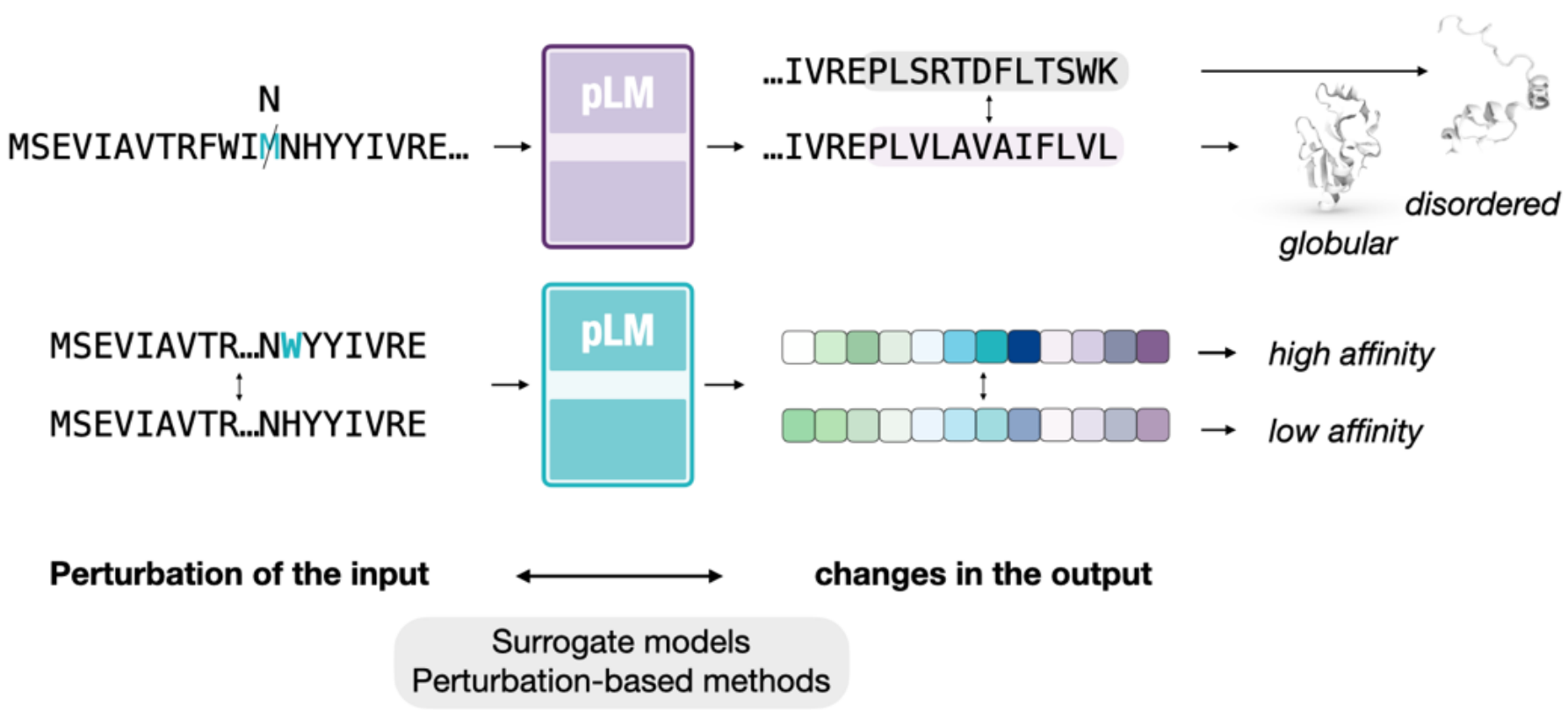

***Figure 5**: Through systematic perturbations of the input and observed changes in the output, the behavior of the model can be probed. This can be used to evaluate the influence of the input features with SHAP, approximate the behavior with interpretable models using LIME, identify sudden changes in the activity landscape with counterfactuals, and test the model's robustness through adversarial attacks.*

An alternative approach to interpret the pLMs' decision process consists of observing changes in the output, by systematically perturbing the input prompt (Fig. 5). Local and global explanations can be inferred from this process, depending on the diversity of the input perturbations. Local explanations focus on understanding individual predictions by perturbing specific inputs, while global explanations examine the model's overall decision-making process by analyzing its responses across multiple inputs. Two of the most widely known techniques in this category are Shapley Additive Explanations (SHAP)[53] and Local Interpretable Model-Agnostic Explanations (LIME)[54]. These automated probing techniques were developed to save time and computational resources compared to brute force testing all possible inputs. SHAP has been used more frequently than LIME on pLMs, but the applications of both methods can be roughly divided into two cases: (1) Embeddings are obtained from a pretrained pLM for transfer learning, where both methods calculate the importance of the embedding features for downstream tasks. SHAP was implemented on tasks such as predicting binding sites[119,120], protein modification sites[121,122], and immunogenicity[123,124] and LIME explained the feature importance for the annotation of antimicrobial peptides[125]. (2) The generated explanations are based on the sequence input and SHAP was used for tasks like the prediction of protein modification sites[126], immunogenicity[127], stability[128], antiviral activity and toxicity[129] and LIME for immune system binding prediction[130].

Counterfactuals and contrastive explanations are instance-based approaches. Counterfactuals detect what minimal (hypothetical) changes in the input would lead to changes in the output,[131] whereas contrastive explanations identify what features (amino acids) led the model to produce one outcome instead of an alternative.[132] Adversarial attacks, which also involve small, intentional input modifications, aim to mislead the model into making incorrect predictions by exploiting its vulnerabilities.[133,134] These probing approaches have been applied in protein research. Previous work has used counterfactual-like reasoning to detect whether structure prediction tools make physically informed decisions when predicting 3D structures.[135–137] ExplainableFold[136] extracts AlphaFold's most impactful amino acids maintaining structural integrity, as well as the most radical or safe substitutions, turning the counterfactual-like analysis into a prediction task. Adversarial attacks are used via prompt injection of jailbreaking[138] to ensure the robustness of LLMs against generating harmful content upon the prompting of malicious perturbations. In the realm of protein research, studies have measured the robustness of the predictions with adversarial attacks,[139,140] and probing could present advantages in other contexts too. In particular, pLMs and genomic language models have the risk of generating sequences of harmful proteins, such as toxins or viral proteins,[141–143] an issue often mitigated by excluding those subsets from the training sets.[110] Another reason is to assess the privacy of the training data, particularly when proprietary or non-public data has been used. In the field of drug discovery, *inference attacks* were used to detect training set leakage, showing that molecules with low similarity to the training set were the most vulnerable, a fact that can be mitigated by representing them as graphs. Outlier molecules with low similarity to the training set were particularly susceptible to identification through probing.[144]

### Potential roles for XAI methods

We also examined the specific purposes for which XAI is employed in pLM research. Adadi et al.[145] previously identified four key needs for XAI: to *justify*, to *control*, to *improve,* and to *discover*. Here, we aimed to explicitly define five distinct *roles* XAI is playing in protein research (**Fig. 6a**), emphasizing its active contribution within workflows to enhance model reliability, efficiency, and versatility, while also enabling novel insights into the underlying data and biological processes.

We found that XAI is most often used as a validation tool to rediscover patterns already described in the literature. In this context, researchers identify patterns that the model is expected to have learned during training. For example, in protein-binding prediction models, one expects amino acids known to interact with a binding partner to receive greater importance, assuming the model has captured the relevant biophysical principles. When XAI methods are applied to LLMs for this purpose, we define this role as the *Evaluator* (**Fig. 6a**). This process does not directly generate new biological knowledge but provides insights into the model's internal reasoning, specifically, whether it has learned what humans expect it to learn. Researchers should, however, remain cautious of human and confirmation bias. Notably, the Evaluator often serves as a prerequisite for other roles, helping to identify model features that subsequent analyses may build on.

In some studies, researchers have gone a step further by demonstrating the generalizability of the extracted patterns, e.g. consistently high attention values for binding residues, and applied this to previously unannotated examples (**Fig. 6a**). This provides new insights into the data itself, and because the model is used for an additional annotation task than the Evaluator, we refer to this role as the *Multitasker*. However, these insights remain confined to patterns already known to exist in nature, which we must predefine. Moreover, even if the model learned such patterns, it remains unclear whether it uses this information in its primary predictive task.

The following two roles focus on the model, and they often rely on insights first revealed by the *Evaluator.* In the *Engineer* role, researchers identify where specific information is encoded, such as in specialized attention heads, and use this to modify the model's architecture or computation pathways while preserving its performance. Examples would be the reduction of the model's computational resources, through intentional structural pruning, shortcutting computations, or optimizing the architecture, while maintaining performance. The *Coach* role aims to change the model output. This can be done without changing the model, by directly up- or down-regulating features to steer predictions, as demonstrated with SAEs[115], or by modifying model weights using insights extracted from XAI analyses. A feature may be considered desirable based on its statistical association with favored outputs or optionally with support from the Evaluator, which can connect it to relevant biological concepts. The Coach role has met success in diffusion architectures. Diffusion models have recently become prominent for both unconditioned and conditioned generation of protein sequences[146]. Because their architecture sometimes contains components in common with pLMs (e.g, encoder modules in continuous sequence diffusion models and attention mechanisms), they have benefited from XAI methods tailored for NLP. In text-to-image diffusion, XAI has been applied through the analysis of cross- and self-attention, counterfactuals generation, concept extraction, training attribution and influence functions, drawing on multiple information sources. In protein research, CMADiff[147] integrates an interpretable BioAligner module to enforce interpretability by design, similar to concept bottleneck models, while Gruver et. al[148] applied gradient-based feature attribution to identify sequence positions most critical for achieving generation objectives.

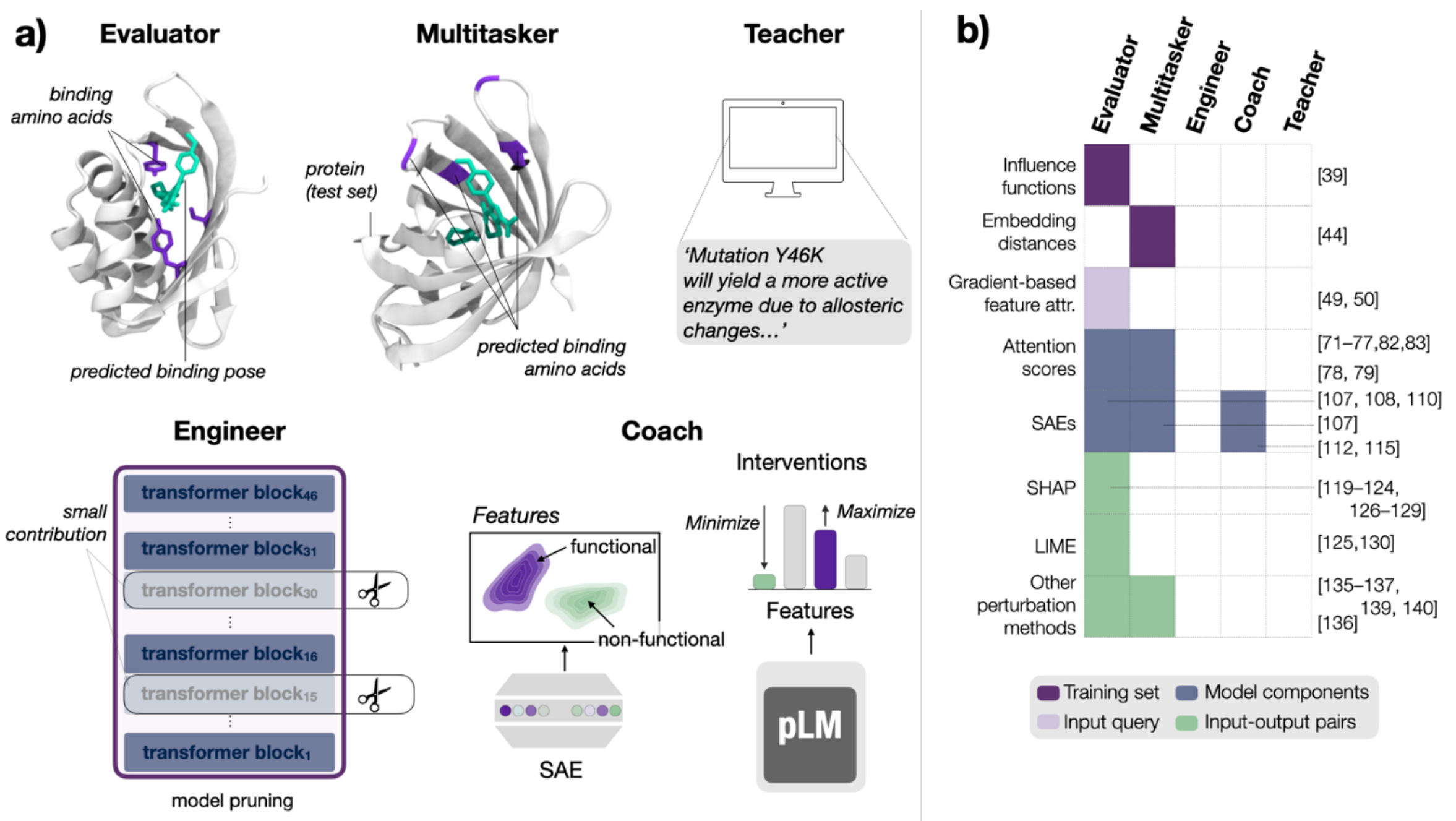

**Figure 6**: **Roles of XAI methods within protein research. (a)** The different identified XAI roles: Evaluator, Multitasker, Teacher, Engineer, and Coach. **(b)** Matrix summarizing various XAI methods and the roles they have been shown to fulfill in pLMs and related methods, color-coded by their information context category. References in the text corresponding to each method are shown in brackets.

Lastly, we define the role of the Teacher for cases where XAI enables the discovery of truly novel and biologically meaningful patterns. This role arises when XAI helps translate learned internal representations into human-understandable concepts without relying on strong human priors. Creating the possibility of learning something unforeseen requires decoupling explanations from human bias; otherwise, we evaluate model-derived insights only through the lens of our current understanding. This contrasts with the Evaluator and Multitasker, where explanations depend directly on human expectations and known biology. It has been hypothesized that pLMs may learn an underlying "protein grammar" or "language of life",[149,150] which is an intriguing idea given the non-intuitive nature of biological sequences. However, because of the statistical nature of predictions and memorization effects, researchers must be careful not to infer incorrect causal relationships from what these models output.[151] For instance, a recent study on chemical language models showed that they often do not infer chemical rules, instead, they rely on statistical correlations in their training data[152]. The same principle may well apply to pLMs: what appears to be a "biologically meaningful" pattern might simply reflect statistical over- or under-representation in the training set, without genuine biological interpretation. Therefore, while realizing the potential of the Teacher remains an open challenge (Box 2), its realization is essential, and it could shed light on fundamental processes such as protein folding or enzyme catalysis. The Teacher represents arguably the most potentially transformative function of XAI, moving beyond improved predictions toward genuine insights into the principles of life.

Overall, XAI has the potential to serve all five roles. While in some cases similar effects could potentially be achieved with other established machine learning concepts, such as parameter reduction via model distillation, XAI-driven approaches place emphasis on explanation first, e.g., the identification of obsolete parts followed by a deliberate reduction of model size in the role of the Engineer. Going forward, it is important for researchers to define the intended purpose of XAI within their deep-learning workflows and to extend its use beyond the evaluation step, making advances toward the Teacher role. We believe these roles are not limited to protein research; they could benefit many areas of life sciences research and deepen our understanding of XAI's possibilities and constraints. This conceptualization represents an essential, though not final, step. As new XAI methods continue to be developed, and new challenges arise, the field will continue to evolve, and additional roles are likely to emerge.

## BOX 2: Toward the Teacher role - when XAI teaches biology

XAI has so far acted as an Evaluator, confirming whether pLMs capture known biological patterns. Using XAI to extract new biological insights from protein AI models, i.e., reaching the Teacher role (Fig. 6a) could deepen our understanding of proteins, but achieving this will require conceptual and methodological advances on several fronts.

A first requirement is that interpretability must be built on robust and faithful foundations. *Faithfulness* refers to the ability of algorithms to produce explanations that reflect the true underlying behavior of the model, instead of giving results that are irrelevant or that fulfill researchers' expectations.[153] This is particularly challenging given the absence of ground truth, which makes it difficult to assess whether explanations genuinely capture model reasoning. The issue is especially pronounced for feature attribution, where different methods often yield conflicting results, raising doubts on their fidelity.[57] Establishing appropriate benchmarks[154–156] and robust evaluation frameworks[56,59,157] as well as strengthening reliability through cross-validating results across methods and information contexts (for example, integrating SHAP with LRP[158] or validating attention analyses with perturbation experiments[78]) will therefore be essential.

*Ante-hoc explainability*, or intrinsic interpretability, refers to models that are transparent by design, such as rule-based systems, decision trees, or linear models. Their strength lies in directly linking decisions to internal representations, but their architectures often reduce flexibility and can limit predictive performance compared to large, black-box deep learning models. Hybrid approaches aim to balance this trade-off by embedding interpretable components into otherwise opaque models. Concept bottleneck pLMs exemplify this direction, as they tie generation to human-understandable intermediate concepts, thereby enhancing interpretability and even enabling controllable generation. In this sense, Karimi *et al.*[159] showed that attention scores alone were insufficient for compound-protein contact prediction, and alleviated the issue by supervising attention, applying structure-aware regularizations and introducing atomic-level "relations" directly in the architecture for intrinsic explainability.

Explainability methods should be tailored to the beneficiaries of the explanation,[160] focusing on human-centered approaches.[161] Explanations designed for model developers (Engineer role), differ from those aimed at experimental biologists or domain experts[162]. Selecting suitable **visualization strategies** is therefore critical.[163,164] While the NLP community already benefits from numerous online tools[45,165–174] and Python packages,[175–177] most remain optimized for text and require adaptation for biological sequences, which are not directly human-readable. In this context, the link between protein structure and function offers a natural bridge: explanations could be projected onto three-dimensional structures[82,178,179] to make them more intuitive (**Fig. 3b**). Multimodal models, like SaProt[21], ProSST[61], ProstT5[23], that incorporate sequence and structural information jointly during training, allow to directly operate on these inputs with feature attribution methods. Alternatively, adding a modality for natural language interfaces, such as ProtChatGPT[180] and Evola[181], could render explanations more accessible. In this sense, multi-agent or self-explaining systems,[182,183–185] (i.e., models that "speak" both sequence and natural language) could offer more intuitive explanations. These models would not only propose a mutation but also articulate why it might increase activity, e.g., by stabilizing a catalytic network or improving packing (**Fig. 6a**), though such approaches risk producing persuasive yet misleading rationales, a phenomenon known as *scheming*.[186]

Equally important is the interplay between XAI insights, model performance, and data quality. Only when biologically relevant patterns are present in the **data**, consistently learned by the model, and faithfully revealed by XAI can interpretable discovery be achieved. In protein science, it is important to bear in mind that models cannot by themselves induce physical and chemical laws unless present in the data or are explicitly integrated during training, such as through physics-aware training paradigms.[187] In this sense, Li et al.[188] recently illustrated how an XAI system could rediscover Kepler's empirical laws of planetary motion from historical astronomical data and then, through symbolic reasoning, obtained Newton's law of universal gravitation. Their framework demonstrated how AI can make faithful predictions based on the data (Kepler's law), but still relied on human scientists to interpret and assign meaning to the discovered equations (Newton's laws). Another recent work applied to the same data also found LLMs excel at their training task (predicting orbital trajectories) but fail to adapt to apply Newtonian mechanics when adapted to new physics tasks.[189] Similarly, in protein modeling, an SAE feature or an attention map remains a mathematical explanation until a human researcher recognizes it as reflecting, for example, an electrostatic network or catalytic motif - a process aligned with the Evaluator or Multitasker roles and dependent on prior knowledge. Achieving the Teacher role will therefore depend not only on trustworthy algorithms but also on interpretation and wet-lab validation. While routine in protein research, wet-lab validation of XAI-derived insights has not been explored to date, and it will be crucial to advance XAI protein research.

Lastly, just as visualization approaches must be adapted to make explanations human-interpretable, the XAI methods themselves that were developed for NLP or computer vision may offer only limited advances toward the Teacher role, because biological sequences are not inherently human-readable. Developing interdisciplinary methods designed with biological sequences in mind could therefore be essential moving forward. Progress may also benefit from cross-disciplinary exchange with fields facing similar challenges in interpreting non-intuitive representations, such as in the success uncovering chess strategies from AlphaZero[190] or reconstructing ancient languages.[191,192] Advancing all these efforts could pave the way toward the Teacher role, where explanations provide genuinely novel biological knowledge.

## Conclusions

pLMs are now central to tasks ranging from structure prediction and variant scoring to sequence generation, yet their internal reasoning is often opaque. In this review, we organized explainability efforts across four information contexts: training data, input prompt, model components, and input–output pairs, and mapped concrete techniques to each. From the literature, we distilled five complementary roles for XAI in protein modeling: Evaluator (validate that models recover known biology), Multitasker (transfer extracted patterns to new annotation tasks), Engineer (prune/reshape architectures using localized signals), Coach (steer training or activations, e.g. via SAEs, to guide outputs), and Teacher (the aspirational role: extract novel, mechanistically credible insights about protein biology). While most prior work has focused on the Evaluator role, achieving genuine biological discovery will require working across several axes. Achieving the Teacher role will be very challenging and, even without guaranteed success, will require robust benchmarks, transparent evaluation, human-centered interfaces, and comprehensive experimental testing. With these pieces in place, XAI may move from post-hoc diagnostics to a design and discovery partner, improving the reliability of predictions, enabling controllable generation, and ultimately helping to uncover principles that govern protein evolution, folding, and function.

## Acknowledgements

We thank Santiago Villalba, Peter Hartog, Alex Vicente, Marcel Hiltscher, Mateusz Iwan and Gerard Boxó for helpful discussions and feedback on this work. We thank the reviewers of the manuscript for the helpful and constructive feedback. This project has received funding from the European Union's Horizon Europe under grant agreement No 101120466 (MSCA-DN supporting A.H.). N.F. acknowledges support from a Ramón y Cajal contract RYC2021-034367-I funded by MCIN/AEI/10.13039/501100011033 and by the European Union NextGenerationEU/PRTR. This work is supported by an grant ATHENA (ERC-ST-2024, Grant agreement 101165231). Views and opinions expressed are, however, those of the author(s) only and do not necessarily reflect those of the European Union. Neither the European Union nor the granting authority can be held responsible for them.

## Conflict of interest

The authors declare no conflict of interest.